\documentclass[a4paper,12pt]{article}
\usepackage{amsmath}
\usepackage{amssymb}
\usepackage{latexsym}
\newcommand{\be}{\begin{eqnarray}}
\newcommand{\ee}{\end{eqnarray}}
\newcommand{\bdm}{\begin{displaymath}}
\newcommand{\edm}{\end{displaymath}}
\begin{document}
\numberwithin{equation}{section}
\title{\Large\textbf{Quantum Cosmology}}
\author{\textbf{T. Christodoulakis}\thanks{e-mail:
tchris@cc.uoa.gr}}
\date{}
\maketitle
\begin{center}
\textit{University of Athens, Physics Department\\
Nuclear \& Particle Physics Section\\
Panepistimioupolis, Ilisia GR 157--71, Athens, Hellas}
\end{center}
\vspace{1cm} \numberwithin{equation}{section}
\begin{abstract}
The problems encountered in trying to quantize the various
cosmological models, are brought forward by means of a concrete
example. The Automorphism groups are revealed as the key element
through which G.C.T.'s can be used for a general treatment of
these problems. At the classical level, the time dependent
automorphisms lead to significant simplifications of the line
element for the generic spatially homogeneous geometry, without
loss of generality. At the quantum level, the ''frozen''
automorphisms entail an important reduction of the configuration
space --spanned by the 6 components of the scale factor matrix--
on which the Wheeler-DeWitt equation, is to be based.\\
In this spirit the canonical quantization of the most general
minisuperspace actions --i.e. with all six scale factor as well as
the lapse function and the shift vector present-- describing the
vacuum type II, I geometries, is considered. The reduction to the
corresponding physical degrees of freedom is achieved through the
usage of the linear constraints as well as the quantum version of
the entire set of all classical integrals of motion.
\end{abstract}
\newpage
\section{\it{Introduction}}
Since the conception by Einstein of General Relativity Theory, a
great many efforts have been devoted by many scientists to the
construction of a consistent quantum theory of gravity. These
efforts can de divided into two main approaches:
\begin{itemize}
\item[(a)] perturbative, in which one splits the metric into a background
(kinematical) part and a dynamical one:
$g_{\mu\nu}=\eta_{\mu\nu}+h_{\mu\nu}$ and tries to quantize $h_{\mu\nu}$.
The only conclusive results existing, are
that the theory thus obtained is highly nonrenormalizable \cite{goroff}.
\item[(b)] non perturbative, in which one tries to keep the twofold role
of the metric (kinematical and dynamical) intact.
A hallmark in this direction is canonical quantization.
\end{itemize}
In trying to implement this scheme for gravity, one faces the problem
of quantizing a constrained system.
The main steps one has to follow are:
\begin{itemize}
\item[(i)] define the basic operators $\widehat{g}_{\mu\nu}$
and $\widehat{\pi}^{\mu\nu}$ and the canonical
commutation relation they satisfy.
\item[(ii)] define quantum operators $\widehat{H}_{\mu}$ whose
classical counterparts are the constraint functions
$H_{\mu}$.
\item[(iii)] define the quantum states $\Psi[g]$ as the common null
eigenvector of $\widehat{H}_{\mu}$, i.e. these
satisfying $\widehat{H}_{\mu}\Psi[g]=0$. (As a consequence, one
has to check that $\widehat{H}_{\mu}$, form a
closed algebra under the basic CCR.)
\item[(iv)] find the states and define the inner product in the space of
these states.
\end{itemize}
It is fair to say that the full program has not yet been carried out,
although partial steps have been made \cite{zanelli}.

In the absence of a full solution to the problem, people have turned to
what is generally known as quantum cosmology.
This is an approximation to quantum gravity in which one freezes out all but a finite number of degrees of freedom,
and quantizes the rest. In this way one is left with a much more manageable problem that is essentially quantum mechanics
with constraints. Over the years, many models have appeared in the
literature \cite{halliwell}. In most of them,
the minisuperspace is flat and the gravitational field is represented by
no more than three degrees of freedom
(generically the three scale factors of some anisotropic Bianchi Type
model \cite{amsterdamski}).

In order for the article to be as self consistent as possible, we
include in section 2, a short introduction to the theory of
constrained systems and in section 3, the Kantowski-Sachs model is
treated both at the classical and the quantum level --as an
interdisciplinary example. In section 4, the importance of the
Automorphism group is uncovered and the quantization of the most
general Type II, I Vacuum Bianchi Cosmologies, is exhibited.
\section{\it{Elements of Constrained Dynamics}}
\subsection{\it{Introduction}}
In these short notes, we present the elements of the general
methods and some techniques of the Constrained Dynamics. It is
about a powerful mathematical theory (a method, more or less)
--primarily developed by P. A. M. Dirac. The scope of it, is to
describe singular (the definition is to be presented at the next
section) physical systems, using a generalization of the
Hamiltonian or the Lagrangian formalism. This theory, is
applicable both for discrete (i.e. finite degrees of freedom) and
continua (i.e. infinite degrees of freedom) systems.

For the sake of simplicity, the Hamiltonian point view of a physical
system is adopted, and the discussion will be restricted on discrete systems.
A basic bibliography, at which the interested reader is strongly
suggested to consult,  is quoted at the end of these notes.
Also, the treatment follows reference \cite{dirac}.
\subsection{\it{The Hamiltonian Approach}}
Suppose a discrete physical system, whose action integral is: \be
\label{action} \mathcal{A}=\int{L}dt \ee The dynamical
coordinates, are denoted by $q^{i}$, with $i \in [1,\ldots,N]$ The
Lagrangian is a function of the
coordinates and the velocities, i.e. $L=L(q^{i},\dot{q}^{i})$.\\
A note is pertinent at this point. If one demands the action
integral (\ref{action}), to be scalar under general coordinate
transformations (G.C.T.), then he can be sure that the content of
the theory to be deduced, will be relativistically covariant even
though the form of the deduced equations will not be manifestly
covariant, on account of the appearance of one particular time in
a dominant place in the theory (i.e. the time variable $t$
occurring already, as soon as one introduces the generalized
velocities, in consequently the Lagrangian, and finally the
Lagrange
transformation, in order to pass from the Lagrangian, to the Hamiltonian).\\
Variation of the action integral, gives the Euler-Lagrange equations
of motion:
\be \label{eulerlagrange}
\frac{d}{dt}\left(\frac{\partial L}{\partial \dot{q}^{i}}\right)=\frac{\partial L}{\partial q^{i}},~~~i \in [1,\ldots,N]
\ee

In order to go over to the Hamiltonian formalism, the momentum
variables $p_{i}$, are introduced through: \be
p_{i}=\frac{\partial L}{\partial \dot{q}^{i}},~~~\forall~~~ i \ee
In the usual dynamical theories, a very restricting assumption is
made; that all momenta are independent functions of the
velocities, or --in view of the inverse map theorem for a function
of many variables-- that the following (Hessian) determinant: \be
|H_{ij}|=|\frac{\partial^{2}L}{\partial \dot{q}^{i}\dot{q}^{j}}|
\ee is not zero in the whole domain of its definition. If this is
the case, then the theorem guarantees the validity of the
assumption, permits to use the Legendre transformation, and the
corresponding physical system is called \emph{Regular}. If this is
not the case (i.e. some momenta, are not independent functions od
the velocities), then there must exist some (say $M$) independent
relations of the type: \be \label{primaryconstraints}
\phi_{m}(q,p)=0,~~~m \in [1,\ldots,M] \ee which are called
\emph{Primary Constraints}. The corresponding physical systems,
are characterized as \emph{Singular}.

Variation of the quantity $p_{i}\dot{q}^{i}-L$ (the Einstein's summation
convention is in use), results in:
\be
\delta\left(p_{i}\dot{q}^{i}-L\right)=\ldots=\left(\delta p_{i}\right)\dot{q}^{i}-\left(\frac{\partial L}{\partial q^{i}}\right)
\delta q^{i}
\ee
by virtue of (\ref{eulerlagrange}). One can see that this variation,
involves variations of the $q$'s and the $p$'s. So, the
quantity under discussion does not involve variation of the velocities
and thus can be expressed in terms of the
$q$'s and the $p$'s, only. This is the Hamiltonian. It must be laid stress
on the fact that the variations, must respect the
restrictions (\ref{primaryconstraints}), i.e. to preserve them --if they are considered as conditions (see, e.g.
C. Carath\'{e}odory, `'Calculus of Variations and Partial Differential
Equations of the First Order`',
AMS Chelsea  (1989)).

Obviously, the Hamiltonian is not uniquely determined for, zero quantities can be
added to it. This means that the following:
\be \label{totalhamiltonian}
H_{T}=H+u^{m}\phi_{m}
\ee
where $u^{m}$'s are arbitrary coefficients in the phase space (including the time variable), is a valid Hamiltonian too.
Variation of (\ref{totalhamiltonian}) results in:
\be \label{equationsofmotion}
\begin{array}{l}
\dot{q}^{i}=\frac{\partial H}{\partial p_{i}}+u^{m}\frac{\partial \phi}{\partial p_{i}}+\textrm{term that vanishes as (\ref{primaryconstraints})}\\
\dot{p}_{i}=-\frac{\partial H}{\partial q_{i}}-u^{m}\frac{\partial \phi}{\partial q_{i}}-\textrm{term that vanishes as (\ref{primaryconstraints})}
\end{array}
\ee
These are the Hamiltonian equations of motion for the system under consideration. This scheme, reflects the previous
observation about variations under which, conditions must be preserved.

In order to proceed, a generalization of the Poisson Brackets must be introduced. This is done as follows:\\
Let $f$, $g$, $h$ be quantities on a space, endowed with a linear
map $\{~,~\}$ such that: \be \label{algebra}
\begin{array}{llll}
\{f,g\}+\{g,f\}=0 & \textrm{Antisymmetry}\\
\{f+g,h\}=\{f,h\}+\{g,h\} &  \textrm{Linearity}\\
\{fg,h\}=f\{g,h\}+\{f,h\}g &  \textrm{Product Law}\\
\{f,\{g,h\}\}+\{g,\{h,f\}\}+\{h,\{f,g\}\}=0 & \textrm{Jacobi Identity}
\end{array}
\ee If the space is the phase space, then these Generalized
Poisson Brackets, reduce to the usual ones: \be \label{definition}
\{f,g\}=\frac{\partial f}{\partial q^{i}}\frac{\partial
g}{\partial p_{i}}-\frac{\partial g}{\partial q^{i}}\frac{\partial
f}{\partial p_{i}} \ee otherwise are subject to the previous
algebra --only.

For a dynamical variable --say $g$, one can find --with the usage
of: \be \dot{g}=\frac{\partial g}{\partial
q^{i}}\dot{q}^{i}+\frac{\partial g}{\partial p_{i}}\dot{p}_{i} \ee
and of (\ref{equationsofmotion}), as well as the generalized
Poisson Bracket Algebra (\ref{algebra}): \be  \label{motionofg}
\dot{g}\approx \{g,H_{T}\} \ee The symbol $\approx$ is the
\emph{Weak Equality} symbol and stands for the following rule
(deduced from a thorough analysis of the previous procedure):\\
\emph{A constraint, must not be used before all the Generalized Poisson Brackets, are calculated formally (i.e. only
with the usage of the algebra (\ref{algebra}) and the usual definition (\ref{definition}) --when the last is applicable)}.
This rule, is encoded as:
\be
\phi_{m}(q,p)\approx 0,~~~m \in [1,\ldots,M]
\ee
In the previous procedure, the position of that rule, reflects the need to manipulate the $u^{m}$'s, which may depend
on $t$ only --since they are unknown coefficients, the definition (\ref{definition}) can not be used.

If the dynamical variable $g$ is any one of the constraints, then (\ref{primaryconstraints}) declare the preservation
of zero. Thus, consistency conditions, are deduced:
\be \label{consistencyconditions}
\{\phi_{m'},H\}+u^{m}\{\phi_{m'},\phi_{m}\}=0
\ee
There are three possibilities:
\begin{itemize}
\item[$CC_{1}$] Relations (\ref{consistencyconditions}) lead to identities --maybe, with the help of
(\ref{primaryconstraints}).
\item[$CC_{2}$] Relations (\ref{consistencyconditions}) lead to equations independent of the $u$'s.
These must also be regarded as constraints. They are called
\emph{Secondary}, but must be treated on the same footing as the
primary ones.
\item[$CC_{3}$] Relations (\ref{consistencyconditions}) impose conditions on the $u$'s.
\end{itemize}
The above procedure must be applied to all secondary constraints.
Again, the possible cases will be the previous three. The new
constraints which may turn up are called secondary too. The
procedure is applied for once more and so on. At the end, one will
have a number of constraints (primary plus secondary) --say
$\mathcal{J}$-- and a number of conditions on the $u$'s. A
detailed analysis of the set of these conditions, shows that: \be
\label{us} u^{m}=U^{m}(q,p)+\mathcal{V}^{a}(t)V^{m}_{a}(q,p) \ee
where $V^{m}_{a}(q,p)$ are the $a$ (in number) independent
solutions of the homogeneous systems: \bdm
V^{m}_{a}(q,p)\{\phi_{m''},\phi_{m}\}=0 \edm The functions
$\mathcal{V}^{a}(t)$ are related to the gauge freedom of the
physical system.

Some terminology is needed at this point.\\
A dynamical variable $R$, is said to be \emph{First Class}, if it
has zero Poisson Bracket, with \underline{all} the constraints:
\be \{R,\phi_{n}\}=0,~~~n \in [1,\ldots,\mathcal{J}] \ee where
$\mathcal{J}$ is the total number of constraints --i.e. primary
plus all the secondary ones. It is sufficient for these
conditions, to hold weakly --since, by definition, the $\phi$'s
are the only independent quantities that vanish weakly. Otherwise,
the variable $R$, is said to be \emph{Second Class}. If $R$ is
First Class, then the quantity $\{R,\phi_{n}\}$ is strongly equal
to some linear combination of the $\phi$'s.
The following relative theorem (with a trivial proof) holds:\\
\emph{''The Poisson Bracket of two First Class quantities, is also First Class''.}

Using the result (\ref{us}) the Hamiltonian
(\ref{totalhamiltonian}), which is called \emph{Total
Hamiltonian}, is written: \be
H_{T}=H+U^{m}\phi_{m}+\mathcal{V}^{a}V^{m}_{a}\phi_{m}\equiv H'
+\mathcal{V}^{a}\phi_{a} \ee with obvious associations. It can be
proved that $H'$ and $\phi_{a}$, are first class quantities.
With this splitting and the relation (\ref{motionofg}) for a dynamical variable $g$, it can be deduced that:\\
\emph{The First Class Primary Constraints $\phi_{a}$, are the
generating functions (i.e. the quantities $\{g,\phi_{a}\}$) of
infinitesimal Contact Transformations; i.e. of transformations
which lead to changes in the $q$'s and the $p$'s which do not
affect the physical state of the system}.

Successive application of two contact transformations generated by
two given First Class Primary Constraints and taking into account
the order, leads --for the sake of consistency-- to a new
generating function: $\{g,\{\phi_{a},\phi_{a'}\}\}$. Thus one can
see that First Class Secondary Constraints, which may turn up from
$\{\phi_{a},\phi_{a'}\}$, can also serve as generating functions
of infinitesimal Contact Transformations. Possibly, another way to
produce  First Class Secondary Constraints, is the First Class
quantity $\{H',\phi_{a}\}$. Since no one has found an example of a
First Class Secondary Constraint, which affects the physical state
when used as generating function, the conclusion is that all First
Class quantities, are  generating functions of infinitesimal
Contact Transformations. Thus, the total Hamiltonian should be
replaced by the \emph{Extended Hamiltonian} $H_{E}$, defined as:
\be H_{E}=H_{T}+\mathcal{U}^{a''}\phi_{a''} \ee where the
$\phi_{a''}$'s are those First Class Secondary Constraints, which
are not already included in $H_{T}$. Finally, the equation of
motion for a dynamical variable $g$ (\ref{motionofg}) is altered:
\be \dot{g}\approx\{g,H_{E}\} \ee
\subsection{\it{Quantization of Constrained Systems}}
\subsubsection{\it{No Second Class Constraints are Present}}
The quantization of a classical physical system, whose Lagrangian, gives first class constraints only, is made in three steps:
\begin{itemize}
\item[$S_{1}$] The dynamical coordinates $q$'s and momenta $p$'s, are turned into Hermitian Operators
$\widehat{q}$~'s and $\widehat{p}$~'s, satisfying
the basic commutative algebra: $[\widehat{q}^{i},\widehat{p}_{j}]=i\delta^{i}_{j}$.
\item[$S_{2}$] A kind of a  Schr\"{o}dinger equation, is set up.
\item[$S_{3}$] Any dynamical function, become Hermitian Operator --provided that the ordering problem is
somehow solved.
\end{itemize}
Obviously, the constraints --being functions on the phase space-- are subject to the $S_{3}$ rule. Dirac, proposed
that when the constraints are turned into operators, they must annihilate the wave function $\Psi$:
\be \label{annihilation}
\widehat{\phi}_{i}\Psi=0,~~~\forall~~~i
\ee
Successive application of two such  given conditions and taking into account the order, for sake of consistency, results in:
\be \label{quantumconditions}
[\widehat{\phi}_{i},\widehat{\phi}_{j}]\Psi=0
\ee
In order for operational conditions (\ref{quantumconditions}) not to give new ones on $\Psi$, one demands:
\be
[\widehat{\phi}_{i},\widehat{\phi}_{j}]=C^{k}_{ij},\widehat{\phi}_{k}
\ee
If it is possible for such an algebra to be deduced, then no new operational conditions on $\Psi$ are found and the system
is consistent. If this is not the case, the new conditions must be taken into account and along with the initial ones,
must give closed algebra, otherwise the procedure must be continued until a closed algebra is found.
The discussion  does not end here. Consistency between the operational conditions (\ref{annihilation}) and the
Schr\"{o}dinger equation, is pertinent as well. This lead to:
\be
[\widehat{\phi}_{i},\widehat{H}]\Psi=0
\ee
and consistency know, reads:
\be
[\widehat{\phi}_{i},\widehat{H}]=D^{k}_{i},\widehat{\phi}_{k}
\ee
\subsubsection{\it{Second Class Constraints are Present}}
Suppose we have a classical physical system, whose Lagrangian,
gives second class constraints. Any set of constraints, can be
replaced by a corresponding set of independent linear combinations
of them. It is thus, in principle, possible to make arrangement
such that the final set of constraints, contains as much first
class constraints as possible. Using the remaining --say $S$ in
number-- second class constraints, the following matrix is
defined: \be \Delta_{ij}=[\chi_{i},\chi_{j}],~~~(i,j) \in
[1,\ldots,S] \ee
where $,\chi$'s are the remaining (in classical form)  second class constraints. A theorem can be proved :\\
\emph{''The determinant of this matrix does not vanish, not even weakly''.}

Since the determinant of $\Delta$ is non zero, there is the inverse of this matrix; say $\Delta^{-1}$.
Dirac, proposed a new kind of Poisson Bracket, the $\{~,~\}_{D}$:
\be
\{~,~\}_{D}=\{~,~\}-\sum^{S}_{i=1}\sum^{S}_{j=1}\{~,\chi_{i}\}\Delta^{-1}_{ij}\{\chi_{j},~\}
\ee
These Brackets, are antisymmetric, linear in their arguments, obey the product law and the Jacobi identity.
It holds that:
\be
\{g,H_{E}\}_{D}\approx \{g,H_{E}\}
\ee
because terms like $\{\chi_{i},H_{E}\}$, with $H_{E}$ being first class, vanish weakly.
Thus:
\be
\dot{g}\approx\{g,H_{E}\}_{D}
\ee
But:
\be
\{\xi,\chi_{s}\}_{D}=\ldots=0
\ee
if $\xi$ is any of the $q$'s or the $p$'s. Thus, at the classical level, one may put the second class constraints equal to
zero, before calculating the new Poisson Brackets. That means that:
\begin{itemize}
\item[$M_{1}$] The equations $\chi=0$ may be considered as strong equations.
\item[$M_{2}$] One, must ignore the corresponding degrees of freedom and
\item[$M_{3}$] quantize the rest, according to the general rules, given in the previous section.
\end{itemize}
\newpage
\section{A Pedagogical Example:\\
The Kantowski-Sachs Model} The purpose of the present section is
twofold:
\begin{itemize}
\item to illustrate --at the classical level-- an application of
Dirac's method for constrained systems.
\item to present, in an easy manner, the problems rased by the
quantization of such systems.
\end{itemize}
The example chosen, is that of Kantowski-Sachs reduced Lagrangian
--i.e. of a vacuum cosmological model; thus the interdisciplinary
character of the section, emerges.
\subsection{The Classical Case}
Consider, the Kantowski-Sachs model (described in
\cite{kantowski}), characterized by the line element: \be
ds^{2}=-N^{2}(t)dt^{2}+a^{2}(t)dr^{2}+b^{2}(t)d\theta^{2}+b^{2}(t)sin^{2}(\theta)d\phi^{2}
\ee The corresponding Eintein's Field Equations, are:

\be
G_{00}=-\left(\frac{N(t)}{b(t)}\right)^{2}-2\frac{a'(t)b'(t)}{a(t)b(t)}
-\left(\frac{b'(t)}{b(t)}\right)^{2} \ee \be
G_{11}=-\left(\frac{a(t)}{b(t)}\right)^{2}+\left(\frac{a(t)b'(t)}{N(t)b(t)}\right)^{2}
-2\frac{a(t)^{2}b'(t)N'(t)}{b(t)N(t)^{3}}+2\frac{a(t)^{2}b''(t)}{b(t)N(t)^{2}}\ee
\be
G_{22}=\frac{b(t)a'(t)b'(t)}{a(t)N(t)^{2}}-\frac{a'(t)b(t)^{2}N'(t)}{N(t)^{3}a(t)}
-\frac{b(t)b'(t)N'(t)}{N(t)^{3}}+\frac{b(t)^{2}a''(t)}{a(t)N(t)^{2}}+
\frac{b(t)b''(t)}{N(t)^{2}}\ee \be G_{33}=sin(\theta)^{2}G_{22}
\ee

The first of these $(G_{00})$, is the quadratic constraint
equation i.e. its time derivative vanishes --by virtue of the
other two $(G_{11}, G_{22})$. This is a ''peculiarity'' of
Einstein's system, and reflects the time reparametrization
invariance $t\rightarrow \widetilde{t}=f(t)$. Under such a
transformation, $a(t)$ and $b(t)$ change as scalars
$(a(t)=\widetilde{a}(\widetilde{t}), \textrm{ditto the~} b(t))$
while $N(t)$, changes as density
$(\widetilde{N}(\widetilde{t})d\widetilde{t}=N(t)dt)$, revealing
its nature, as a Lagrange multiplier.

It must be brought to the reader's notice that the above set of
equations $(G_{\mu\nu})$, can be obtained from the following
action principle: \be \mathcal{S}=\int Ldt=\int \left(
-\frac{a(t)\dot{b}^{2}(t)+2b(t)\dot{a}(t)\dot{b}(t)}{2N(t)}+\frac{N(t)a(t)}{2}
\right)dt\ee --where $a(t)$, $b(t)$ and $N(t)$, are the three
degrees of freedom (the $q$'s)-- which has the above mentioned
reparametrization invariance.

The momenta are:
\begin{subequations}
\be \label{momenta}
p_{a}=\frac{\partial L}{\partial \dot{a}(t)} & =  &-\frac{b(t)\dot{b}(t)}{N(t)}\\
p_{b}=\frac{\partial L}{\partial \dot{b}(t)} &  =  & -\frac{\dot{a}(t)b(t)+a(t)\dot{b}(t)}{N(t)}\\
p_{N}= \frac{\partial L}{\partial \dot{N}(t)} &  = & 0 \ee
\end{subequations}
From the third of (\ref{momenta}), one can see that there is one
primary constraint: \be p_{N}\approx 0 \ee

The total Hamiltonian is: \be H_{T}=H+u(t)p_{N} \ee where: \be
H=p_{a}\dot{a}(t)+p_{b}\dot{b}(t)-L=N(t)\Omega(t) \ee with: \be
\label{omega}
\Omega(t)\equiv-\frac{a(t)}{2}-\frac{p_{a}p_{b}}{b(t)}+\frac{a(t)p^{2}_{a}}{2b^{2}(t)}
\ee

The consistency condition (\ref{consistencyconditions}) applied to:
\begin{itemize}
\item[$A_{1}$] the constraint $p_{N}\approx 0$, gives one secondary constraint:
\be \label{chiconstraint}
\chi\equiv\{p_{N},H\}=\{p_{N},N(t)\Omega(t)\}=-\Omega(t)\approx 0
\ee
A straightforward calculation, results in:
\be \label{chipn}
\{\chi,p_{N}\}=0
\ee
\item[$A_{2}$] the previously deduced secondary constraint $\chi\approx 0$, gives --by virtue of (\ref{omega}),
(\ref{chiconstraint}) and (\ref{chipn})--  no further constraints,
since it is identically satisfied ($CC_{1}$ case): \be
\{\chi,H\}+u(t)\{\chi,p_{N}\}=0 \ee
\end{itemize}

The Poisson Bracket (\ref{chipn}) also declares that both $p_{N}$ and $\chi$, are first class quantities.

Finally, the equations of motion are: \be \dot{a}(t)\approx
\{a(t),H_{T}\} \ee \be \dot{p_{a}}\approx \{p_{a},H_{T}\} \ee \be
\dot{b}(t)\approx \{b(t),H_{T}\} \ee \be \dot{p_{b}}\approx
\{p_{b},H_{T}\} \ee \be \label{multiplier} \dot{N}(t)\approx
\{N(t),H_{T}\} \ee \be \label{pn} \dot{p_{N}}\approx
\{p_{N},H_{T}\} \ee The first four equations constitute the usual
set of the Euler-Lagrange equations for the $a(t)$ and $b(t)$,
degrees of freedom. Equation (\ref{multiplier}), results in the
gauge freedom related to $N(t)$ since --according to this
equation-- $\dot{N}(t)=u(t)$, i.e. an arbitrary function of time,
while equation (\ref{pn}) is trivially satisfied, in view of
(\ref{chiconstraint}).

Finally a remark concerning the existence of shift terms of the
form $N_{i}(x^{j},t)dx^{i}dt$ --$x^{j}$ stand for $(r, \theta,
\phi)$-- where $N_{i}(x,t)\equiv N_{a}(t)\sigma^{a}_{i}(x)$: their
existence entails constraint equations $(G_{0i})$ --again
preserved in time, by virtue of the $(G_{ij})$ equations-- which
reflect the space reparametrization invariance $x^{i}\rightarrow
\widetilde{x^{i}}(x^{j},t)$. Along with the existence of these
shift terms, a change in the spatial part of the line element, is
induced.

\subsection{The Quantum Case}

In trying to quantize the previously described constraint
Hamiltonian system, various problems, arise
\cite{dirac,sundermeyer}.

In the canonical approach \cite{zanelli} --and references
therein--, the Schr\"{o}dinger representation, is most frequently
adopted. Applied to our example, this entails the step:
\be\begin{array}{ll} a\rightarrow \widehat{a}=a\\
b\rightarrow \widehat{b}=b\\
N\rightarrow \widehat{N}=N\\
p_{a}\rightarrow \widehat{p}_{a}=-i\frac{\partial}{\partial a}\\
p_{b}\rightarrow \widehat{p}_{b}=-i\frac{\partial}{\partial
b}\\
p_{N}\rightarrow \widehat{p}_{N}=-i\frac{\partial}{\partial N}
\end{array}
\ee

When trying to implement Dirac's proposal (steps $S_{1}$, $S_{2}$
of the section 2.3.1) we came across the factor ordering problem
(see e.g. T. Christodoulakis, J. Zanelli, Nuovo Cimento B
\textbf{93} (1986) 1). Its resolution is achieved via the recipe
that the kinetic term must be realized as the conformal Laplacian.
This is due to the covariance in the change
$\widetilde{N}(\widetilde{t})=N(t)f(a(t),b(t))$ --with the
understanding that $f(a(t),b(t))$, is identified to an arbitrary
function of time. The conformal Laplacian must be based on the
metric: \be g^{ij}=\left(\begin{array}{cc}
  a/2b^{2} & -1/2b \\
  -1/2b & 0
\end{array}\right) \ee
because of the correspondence principle, since
$H=N(g^{ij}p_{i}p_{j}+V)$, where $p_{1}\equiv p_{a}$, $p_{2}\equiv
p_{b}$ and $V=-a/2$. In two dimensions, the conformal Laplacian,
reduces to the typical one (see next section for details). Thus,
following Dirac's quantization program, we deduce: \be
\widehat{H}_{T}\Psi=0 \ee or: \be \widehat{p}_{N}\Psi=0\ee
(Constraint) and: \be\widehat{H}\Psi=0 \ee (Wheeler-DeWitt
equation). Under the transformation: \be\begin{array}{ll}
a\rightarrow u=b\\
b\rightarrow v=a^{3}b\end{array}\ee the Wheeler-DeWitt equation,
assumes the form: \be 4\frac{\partial^{2}\Psi}{\partial u\partial
v}-\Psi=0 \ee and under a second transformation:
\be\begin{array}{ll}
u(t)\rightarrow X=\frac{u+v}{2}\\
v(t)\rightarrow Y=\frac{u-v}{2}\end{array}\ee the Wheeler-DeWitt
equation, takes the form: \be \frac{\partial^{2}\Psi}{\partial
X^{2}}-\frac{\partial^{2}\Psi}{\partial Y^{2}}-\Psi=0 \ee Now the
previous equation can be solved via the method of separation of
variables, e.g. $\Psi(X,Y)=A(X)B(Y)$; its general solution,
consists of products of Exponentials and/or Trigonometric
functions --depending on the sign and the value of the separation
constant.

Of course, in order to complete the program of quantization, we
need to construct the Hilbert space, i.e. to select a measure. The
problem is open, because there is an infinitude of candidates. If
one invokes some sort of ''naturality'', one could adopt as a
measure the square root of the determinant of the supermetric,
i.e. $2b$ --in our case. This however, causes two unpleasant
drawbacks: the first is that the wave function, is not square
integrable, and the second is the violation of the conformal
covariance.

In the case where shift terms and more spatial metric cross terms,
are present, one would like to know, what features of the above
exhibition, are generic and thus, characterize the general
situation. The answer is given through the consideration of the
automorphism group, which can be considered as the symmetries of
the symmetry group of the $3$-space. Their action entails
considerable simplification, both at the classical and the quantum
level. The spirit of these ideas, is exhibited in the next
sections.
\newpage
\section{\it{Automorphisms in Classical \& Quantum Cosmology}}
\subsection{\it{The Simplification of Einstein's Equations}}
It has long been suspected and/or known, that automorphisms, ought
to play an important role in a unified treatment of this problem.
The first mention, goes back to the first of \cite{6paper}. More
recently, Jantzen, --second of \cite{6paper}-- has used Time
Dependent Automorphism Matrices, as a convenient parametrization
of a general positive definite $3\times 3$ scale factor matrix
$\gamma_{\alpha\beta}(t)$, in terms of a --desired-- diagonal
matrix. His approach, is based on the orthonormal frame bundle
formalism, and the main conclusion is (third of \cite{6paper}, pp.
1138): ''\textit{\ldots the special automorphism matrix group
SAut(G), is the symmetry group of the ordinary differential
equations, satisfied by the metric matrix $\gamma_{\alpha\beta}$,
when no sources are present \ldots}'' Later on, Samuel and
Ashtekar in \cite{7paper}, have seen automorphisms, as a result of
general coordinate transformations. Their spacetime point of view,
has led them, to consider the --so called-- ''Homogeneity
Preserving Diffeomorphisms'', and link them, to topological
considerations.
\subsubsection{Time Dependent Automorphism\\ Inducing Dif\mbox{}feomorphisms}
It is well known that the vacuum Einstein field equations can be
derived from an action principle:
\be
\mathcal{A}=\frac{-1}{16\pi}\int\sqrt{-^{(4)}g}~^{(4)}R~d^{4}x
\ee
(we use geometrized units i.e. $G=c=1$)\\
The standard canonical formalism \cite{8paper} makes use of the
lapse and shift functions appearing in the 4-metric: \be
\label{e2.2} ds^{2}=(N^{i} N_{i}-
N^{2})dt^{2}+2N_{i}dx^{i}dt+g_{ij}dx^{i}dx^{j} \ee From this
line-element the following set of equations obtains, expressed in
terms of the extrinsic curvature: \bdm
K_{ij}=\frac{1}{2N}(N_{i\mid j}+N_{i\mid j}-\frac{\partial g_{ij}
}{\partial t}) \edm
\begin{subequations} \label{e2.3}
\be H_{0}=\sqrt{g}~(K_{ij}K^{ij}-K^{2}+R)=0 \ee \be
H_{i}=2\sqrt{g}~(K^{j}_{i\mid j}-K_{\mid i})=0 \ee \be
\frac{1}{\sqrt{g}}\frac{d}{dt}[\sqrt{g}~(K^{ij}-K g^{ij})]=-N
(R^{ij}-\frac{1}{2}R~
g^{ij})-\frac{N}{2}(K_{kl}K^{kl}-K^{2})g^{ij}\\
+2N(K^{ik}K^{j}_{k}-K~K^{ij})- (N^{\mid ij}-N^{\mid l}_{\mid
l}g^{ij})+[(K^{ij}-K~g^{ij})N^{l}]_{\mid l}\\-N^{i}_{\mid
l}(K^{lj}-K~g^{lj})-N^{j}_{\mid l}(K^{li}-K~g^{li}) \ee
\end{subequations}
This set is equivalent to the ten Einstein's equations.

In cosmology, we are interested in the class of spatially
homogeneous spacetimes, characterized by the existence of an
m-dimensional isometry group of motions $G$, acting transitively
on each surface of simultaneity $\Sigma_{t}$. When m is greater
than 3 and there is no proper invariant subgroup of dimension 3,
the spacetime is of the Kantowski-Sachs type \cite{kantowski} and
will not concern us further. When m equals the dimension of
$\Sigma_{t}$ --which is 3--, there exist 3 basis one-forms
$\sigma^{\alpha}_{i}$ satisfying:
\begin{subequations} \label{e2.4}
\be d\sigma^{\alpha}=C^{\alpha}_{\beta\gamma}~\sigma^{\beta}\wedge
\sigma^{\gamma}\Leftrightarrow \sigma^{\alpha}_{i
,~j}-\sigma^{\alpha}_{j,~i}=2C^{\alpha}_{\beta\gamma}~
\sigma^{\gamma}_{i}~\sigma^{\beta}_{j} \ee where
$C^{\alpha}_{\beta\gamma}$ are the structure constants of
the corresponding isometry group.\\
In this case there are local coordinates $t, ~x^{i}$ such that the
line element in (\ref{e2.2}) assumes the form: \be
ds^{2}&=(N^{\alpha}(t) N_{\alpha}(t)-
N^{2}(t))dt^{2}+2N_{\alpha}(t)\sigma^{\alpha}_{i}(x)dx^{i}dt\\&
+\gamma_{\alpha\beta}\sigma^{\alpha}_{i}(x)\sigma^{\beta}_{j}(x)dx^{i}dx^{j}
\ee
\end{subequations}
Latin indices, are spatial with range from 1 to 3. Greek indices,
number the different basis 1-forms, take values in the same range,
and are lowered and raised by $\gamma_{\alpha\beta}$, and
$\gamma^{\alpha\beta}$ respectively.

A commitment concerting the topology of the $3$-surface, is
pertinent here, especially in view of the fact that we wish to
consider diffeomorphisms \cite{7paper}; we thus assume that $G$ is
simply connected and the $3$-surface $\Sigma_{t}$ can be
identified with $G$, by singling out a point $p$ of $\Sigma_{t}$,
as the identity $e$, of $G$.

If we insert relations (\ref{e2.4}) into equations (\ref{e2.3}),
we get the following set of ordinary differential equations for
the Bianchi-Type spatially homogeneous spacetimes:
\begin{subequations} \label{e2.5}
\be E_{0}\doteq K^{\alpha}_{\beta}~K^{\beta}_{\alpha}-K^{2}+R=0
\ee \be E_{\alpha}\doteq
K^{\mu}_{\alpha}~C^{\epsilon}_{\mu\epsilon}-K^{\mu}_{\epsilon}~C^{\epsilon}_{\alpha\mu}=0
\ee \be E^{\alpha}_{\beta}\doteq
\dot{K}^{\alpha}_{\beta}-NKK^{\alpha}_{\beta}+NR^{\alpha}_{\beta}+
2N^{\rho}(K^{\alpha}_{\nu}~C^{\nu}_{\beta\rho}-K^{\nu}_{\beta}~C^{\alpha}_{\nu\rho})
\ee
\end{subequations}
where $K^{\alpha}_{\beta}=\gamma^{\alpha\rho}K_{\rho\beta}$ and
\be \label{e2.6}
K_{\alpha\beta}=-\frac{1}{2N}(\dot{\gamma}_{\alpha\beta}+2\gamma_{\alpha\nu}C^{\nu}_{\beta\rho}N^{\rho}+2\gamma_{\beta\nu}C^{\nu}_{\alpha\rho}N^{\rho})
\ee \be \label{e2.7}
R_{\alpha\beta}&=C^{\kappa}_{\sigma\tau}C^{\lambda}_{\mu\nu}\gamma_{\alpha\kappa}\gamma_{\beta\lambda}\gamma^{\sigma\nu}\gamma^{\tau\mu}+
2C^{\lambda}_{\alpha\kappa}C^{\kappa}_{\beta\lambda}+2C^{\mu}_{\alpha\kappa}C^{\nu}_{\beta\lambda}\gamma_{\mu\nu}\gamma^{\kappa\lambda}\\
&+2C^{\lambda}_{\beta\kappa}C^{\mu}_{\mu\nu}\gamma_{\alpha\lambda}\gamma^{\kappa\nu}+
2C^{\lambda}_{\alpha\kappa}C^{\mu}_{\mu\nu}\gamma_{\beta\lambda}\gamma^{\kappa\nu}
\ee When $N^{\alpha}=0$, equation (\ref{e2.5}c) reduces to the
form of the equation given in \cite{10paper}. Equation set
(\ref{e2.5}), forms what is known as a --complete-- perfect ideal;
that is, there are no integrability conditions obtained from this
system. So, with the help of (\ref{e2.5}c), (\ref{e2.6}),
(\ref{e2.7}), it can explicitly be shown, that the time
derivatives of (\ref{e2.5}a) and (\ref{e2.5}b) vanish identically.
The calculation is staightforward --although somewhat lengthy. It
makes use of the Jacobi identity
$C^{\alpha}_{\rho\beta}C^{\rho}_{\gamma\delta}+C^{\alpha}_{\rho\delta}C^{\rho}_{\beta\gamma}+
C^{\alpha}_{\rho\gamma}C^{\rho}_{\delta\beta}=0$, and its
contracted form
$C^{\alpha}_{\alpha\beta}C^{\beta}_{\gamma\delta}=0$.

The vanishing of the derivatives of the 4 constrained equations:\\
$E_{0}=0, E_{\alpha}=0$, implies that these equations, are first
integrals of equations (\ref{e2.5}c) --moreover, with vanishing
integration constants. Indeed, algebraically solving
(\ref{e2.5}a), (\ref{e2.5}b) for $N(t), N^{\alpha}(t)$,
respectively and substituting in (\ref{e2.5}c), one finds that in
all --but Type II and III-- Bianchi Types, equations
(\ref{e2.5}c), can be solved for only 2 of the 6 accelerations
$\ddot{\gamma}_{\alpha\beta}$ present. In Type II and III, the
independent accelerations are 3, since $E_{\alpha}$ are not
independent and thus, can be solved for only 2 of  the 3
$N^{\alpha}$'s. But then in both of these cases, a linear
combination of the $N^{a}$'s remains arbitrary, and
counterbalances the extra independent acceleration. Thus, in all
Bianchi Types, 4 arbitrary functions of time enter the general
solution to the set of equations (\ref{e2.5}). Based on the
intuition gained from the full theory, one could expect this fact
to be a reflection of the only known covariance of the theory;
i.e. of the freedom to make arbitrary changes of the time and
space coordinates.

The rest of this section is devoted to the investigation of the
existence, uniqueness, and properties of general coordinate
transformations --containing 4 arbitrary functions of time--,
which on the one hand, must preserve the manifest spatial
homogeneity, of the line element (\ref{e2.4}b), and on the other
hand,
must be symmetries of equations (\ref{e2.5}).\\
As far as time reparametrization is concerned the situation is
pretty clear: If a transformation:
\begin{subequations} \label{e2.8}
\be t\rightarrow \tilde{t}=g(t) \Leftrightarrow t=f(\tilde{t })
\ee is inserted in the line element (\ref{e2.4}b), it is easily
inferred that: \be \gamma_{\alpha\beta}(t)\rightarrow
\gamma_{\alpha\beta}(f(\tilde{t}))\equiv
\tilde{\gamma}_{\alpha\beta}(\tilde{t}) \ee \be N(t)\rightarrow
\pm~ N(f(\tilde{t}))\frac{df(\tilde{t})}{d\tilde{t}}\equiv
\widetilde{N}(\tilde{t})\\
N^{\alpha}(t)\rightarrow
N^{\alpha}(f(\tilde{t}))\frac{df(\tilde{t})}{d\tilde{t}}\equiv
\widetilde{N}^{\alpha}(\tilde{t}) \ee
\end{subequations}
Accordingly, $K^{\alpha}_{\beta}$ transforms under (\ref{e2.8}a)
as a scalar and thus (\ref{e2.5}a), (\ref{e2.5}b) are also scalar
equations while (\ref{e2.5}c) gets multiplied by a factor
$df(\tilde{t})/d\tilde{t}$. Thus, given a particular solution to
equations (\ref{e2.5}), one can always obtain an equivalent
solution, by arbitrarily redefining time. Hence, we understand the
existence of one arbitrary function of time in the general
solution to Einstein's equations (\ref{e2.5}). In order to
understand the presence of the rest 3 arbitrary functions of time
it is natural to turn our attention to the transformations of the
3 spatial coordinates $x^{i}$. To begin with, consider the
transformation: \be \label{e2.9}
\tilde{t}=t\Leftrightarrow& t=\tilde{t}\\
\tilde{x}^{i}=g^{i}(x^{j},t)\Leftrightarrow&
x^{i}=f^{i}(\tilde{x}^{j},\tilde{t}) \ee It is here understood,
that our previous assumption concerning the topology of $G$ and
the identification of $\Sigma_{t}$ with $G$, is valid for all
values of the parameter $t$, for which the transformation is to be
well defined.

Under these transformations, the line element (\ref{e2.4}b)
becomes: \be \label{e2.10}
ds^{2}&=[(N^{\alpha}N_{\alpha}-N^{2})+\frac{\partial
f^{i}}{\partial \tilde{t}}\frac{\partial f^{j}}{\partial \tilde{t}
}~\sigma^{\alpha}_{i}(f)\sigma^{\beta}_{j}(f)\gamma_{\alpha\beta}(\tilde{t})\\
&+2\sigma^{\alpha}_{i}(f)\frac{\partial f^{i}}{\partial
\tilde{t}}N_{\alpha}(\tilde{t})]d\tilde{t}^{2}\\
&+2\sigma^{\alpha}_{i}(x)\frac{\partial x^{i}}{\partial
\tilde{x}^{m}}[N_{\alpha}(\tilde{t})+\sigma^{\beta}_{j}(x)\frac{\partial
x^{j}}{\partial
\tilde{t}}\gamma_{\alpha\beta}(\tilde{t})]d\tilde{x}^{m}d\tilde{t}\\
&+\sigma^{\alpha}_{i}(x)\sigma^{\beta}_{j}(x)\gamma_{\alpha\beta}(\tilde{t})\frac{\partial
x^{i}}{\partial \tilde{x}^{m}}\frac{\partial x^{j}}{\partial
\tilde{x}^{n}}~d\tilde{x}^{m}d\tilde{x}^{n} \ee Since our aim, is
to retain manifest spatial homogeneity of the line element
(\ref{e2.4}b), we have to refer the form of the line element given
in (\ref{e2.10}) to the old basis $\sigma^{\alpha}_{i}(\tilde{x})$
at the new spatial point $\tilde{x}^{i}$. Since
$\sigma^{\alpha}_{i}$ --both at $x^{i}$ and $\tilde{x}^{i}$--, as
well as, $\partial x^{i}/
\partial \tilde{x}^{j}$, are invertible matrices, there
always exists a non-singular matrix $\Lambda
^{\alpha}_{\mu}(\tilde{x},\tilde{t})$ and a triplet
$P^{\alpha}(\tilde{x},\tilde{t})$, such that: \be \label{e2.11}
\sigma^{\alpha}_{i}(x)\frac{\partial x^{i}}{\partial
\tilde{x}^{m}}=&\Lambda^{\alpha}_{\mu}(\tilde{x},\tilde{t})\sigma^{\mu}_{m}(\tilde{x})\\
\sigma^{\alpha}_{i}(x)\frac{\partial x^{i}}{\partial
\tilde{t}}=&P^{\alpha}(\tilde{x},\tilde{t}) \ee The above
relations, must be regarded as definitions, for the matrix
$\Lambda ^{\alpha}_{\mu}$ and the triplet $P^{\alpha}$. With these
identifications the line element (\ref{e2.10}) assumes the form:
\be \label{e2.12} ds^{2}&=[(N^{\alpha}N_{\alpha}-
N^{2})+P^{\alpha}(\tilde{x},\tilde{t})P^{\beta}(\tilde{x},\tilde{t})\gamma_{\alpha\beta}(\tilde{t})
+2P^{\alpha}(\tilde{x},\tilde{t})N_{\alpha}(\tilde{t})]d\tilde{t}^{2}\\
&+2\Lambda^{\alpha}_{\mu}(\tilde{x},\tilde{t})\sigma^{\mu}_{m}(\tilde{x})[N_{\alpha}(\tilde{t})
+P^{\beta}(\tilde{x},\tilde{t})\gamma_{\alpha\beta}(\tilde{t})]d\tilde{x}^{m}d\tilde{t}\\
&+\Lambda^{\alpha}_{\mu}(\tilde{x},\tilde{t})\Lambda^{\beta}_{\nu}(\tilde{x},\tilde{t})
\gamma_{\alpha\beta}(\tilde{t})\sigma^{\mu}_{m}(\tilde{x})\sigma^{\nu}_{n}(\tilde{x})d\tilde{x}^{m}
d\tilde{x}^{n} \ee If, following the spirit of \cite{7paper}, we
wish the transformation (\ref{e2.9}) to be manifest homogeneity
preserving i.e. to have a well defined, non-trivial action on
$\gamma_{\alpha\beta}(t)$, $N(t)$ and $N^{\alpha}(t)$, we must
impose the condition that
$\Lambda^{\alpha}_{\mu}(\tilde{x},\tilde{t})$ and
$P^{\alpha}(\tilde{x},\tilde{t})$ do not depend on the spatial
point $\tilde{x}$, i.e.
$\Lambda^{\alpha}_{\mu}=\Lambda^{\alpha}_{\mu}(\tilde{t})$ and
$P^{\alpha}=P^{\alpha}(\tilde{t})$. Then (\ref{e2.12}) is written
as: \be ds^{2}&=[(N^{\alpha}N_{\alpha}-
N^{2})+P^{\alpha}P^{\beta}\gamma_{\alpha\beta}
+2P^{\alpha}N_{\alpha}]d\tilde{t}^{2}\\
&+2\Lambda^{\alpha}_{\mu}\sigma^{\mu}_{m}(\tilde{x})[N_{\alpha}
+P^{\beta}\gamma_{\alpha\beta}]d\tilde{x}^{m}d\tilde{t}\\
&+\Lambda^{\alpha}_{\mu}\Lambda^{\beta}_{\nu}
\gamma_{\alpha\beta}\sigma^{\mu}_{m}(\tilde{x})\sigma^{\nu}_{n}(\tilde{x})d\tilde{x}^{m}
d\tilde{x}^{n}\Rightarrow\\
ds^{2}&\equiv (\widetilde{N}^{\alpha}\widetilde{N}_{\alpha}-
\widetilde{N}^{2})d\tilde{t}^{2}+2\widetilde{N}_{\alpha}(\tilde{t})\sigma^{\alpha}_{i}(\tilde{x})d\tilde{x}^{i}d\tilde{t}\\
&+\widetilde{\gamma}_{\alpha\beta}(\tilde{t})\sigma^{\alpha}_{i}(\tilde{x})\sigma^{\beta}_{j}(\tilde{x})d\tilde{x}^{i}d\tilde{x}^{j}
\ee with the allocations:
\begin{subequations} \label{e2.14}
\be
\widetilde{\gamma}_{\alpha\beta}=\Lambda^{\mu}_{\alpha}\Lambda^{\nu}_{\beta}\gamma_{\mu\nu}
\ee
\be
\widetilde{N}_{\alpha}=\Lambda^{\beta}_{\alpha}(N_{\beta}+P^{\rho}\gamma_{\rho\beta})~~
and~~thus~~\widetilde{N}^{\alpha}=S^{\alpha}_{\beta}(N^{\beta}+P^{\beta})
\ee \be \widetilde{N}=N \ee
\end{subequations}
(where $S=\Lambda^{-1}$).\\
Of course, the demand that $\Lambda^{\alpha}_{\beta}$ and
$P^{\alpha}$ must not depend on the spatial point $\tilde{x}^{i}$,
changes the character of (\ref{e2.11}), from identities, to the
following set of differential restrictions on the functions
defining the transformation:
\begin{subequations} \label{e2.15}
\be \frac{\partial f^{i}}{\partial
\tilde{x}^{m}}=\sigma^{i}_{\alpha}(f)\Lambda^{\alpha}_{\beta}(\tilde{t})\sigma^{\beta}_{m}(\tilde{x})
\ee \be \frac{\partial f^{i}}{\partial
\tilde{t}}=\sigma^{i}_{\alpha}(f)P^{\alpha}(\tilde{t}) \ee
\end{subequations}
Equations (\ref{e2.15}) constitute a set of first-order highly
non-linear P.D.E.'s for the unknown functions $f^{i}$. The
existence of local solutions to these equations is guaranteed by
Frobenius theorem \cite{11paper} as long as the necessary and
sufficient conditions: \bdm \frac{\partial}{\partial
\tilde{x}^{j}}\Big( \frac{\partial f^{i}}{\partial
\tilde{x}^{m}}\Big)- \frac{\partial}{\partial \tilde{x}^{m}}\Big(
\frac{\partial f^{i}}{\partial \tilde{x}^{j}}\Big)=0 \edm \bdm
\frac{\partial}{\partial \tilde{t}}\Big( \frac{\partial
f^{i}}{\partial \tilde{x}^{m}}\Big)- \frac{\partial}{\partial
\tilde{x}^{m}}\Big( \frac{\partial f^{i}}{\partial
\tilde{t}}\Big)=0 \edm hold. Through (\ref{e2.15}) and repeated
use of (\ref{e2.4}a), these equations reduce respectively to: \be
\label{e2.16}
\Lambda^{\alpha}_{\mu}C^{\mu}_{\beta\gamma}=\Lambda^{\rho}_{\beta}\Lambda^{\sigma}_{\gamma}C^{\alpha}_{\rho\sigma}
\ee \be \label{e2.17}
P^{\mu}C^{\alpha}_{\mu\nu}\Lambda^{\nu}_{\beta}=\frac{1}{2}\dot{\Lambda}^{\alpha}_{\beta}
\ee It is noteworthy that the solutions to (\ref{e2.16}) and
(\ref{e2.17}), --by virtue of (\ref{e2.14})-- form a group, with
composition law: \bdm
(\Lambda_{3})^{\alpha}_{\beta}=(\Lambda_{1})^{\alpha}_{\varrho}(\Lambda_{2})^{\varrho}_{\beta}
\edm \bdm
(P_{3})^{a}=(\Lambda_{1})^{\alpha}_{\beta}(P_{2})^{\beta}+(P_{1})^{a}
\edm where ($\Lambda_{1},~P_{1}$) and ($\Lambda_{2},~P_{2}$), are
two
successive transformations of the form (\ref{e2.14}).\\
Note also, that a constant automorphism is always a solution of
(\ref{e2.16}), (\ref{e2.17}); indeed,
$\Lambda^{a}_{\beta}(t)=\Lambda^{a}_{\beta}$ and $P^{a}(t)=0$
solve these equations. Thus, $\Lambda^{a}_{\beta}$ and $P^{a}=0$
can be regarded as the remaining gauge symmetry, after one has
fully used the arbitrary functions of time, appearing in a
solution $\Lambda^{a}_{\beta}(t)$ and $P^{a}(t)$. Consequently one
can, at first sight, regard all the arbitrary constants
encountered when integrating (\ref{e2.17}), as absorbable in the
shift, since the transformation law for the shift, is then
tensorial. This is certainly true, as long as there is a non zero
initial shift. However, if one has used the independent functions
of time, in order to set the shift zero, then the constants
remaining within $\Lambda^{a}_{\beta}$, are not absorbable. It is
this kind of constants that are explicitly present in `'T.
Christodoulakis, G. Kofinas, E. Korfiatis, G. O. Papadopoulos and
A. Paschos, J. Math. Phys. 42 (2001) 3580-3608, gr-qc/0008050`'.
Where the solutions to (\ref{e2.16}),~(\ref{e2.17}) for all
Bianchi Types are given. A relevant nice discussion,
distinguishing between genuine gauge symmetries (cf. arbitrary
functions of time) and rigid symmetries (cf. arbitrary constants),
is presented in \cite{12paper}. There a different definition of
manifest homogeneity preserving diffeomorphisms --stronger than
the one adopted in this work-- is used, and results in only the
inner automorphisms being allowed to acquire $t$ dependence. In
connection to this, it is interesting to observe that
(\ref{e2.16},~\ref{e2.17}) give essentially the same results:
notice that $2P^{\mu}C^{\alpha}_{\mu\beta}$ is, by definition, the
generator of Inner Automorphisms. Thus there is always a
$\lambda^{\alpha}_{\beta}(t)\equiv
Exp(2P^{\mu}C^{\alpha}_{\mu\beta})$ $\in$ IAut(G) satisfying
(\ref{e2.17}). If we now parameterize the general solution to
(\ref{e2.16},~\ref{e2.17}) by
$\Lambda^{\alpha}_{\beta}(t)=\lambda^{\alpha}_{\varrho}(t)U^{\varrho}_{\beta}(t)$
and substitute in these relations, we deduce that the matrix $U$
is a constant automorphism. This analysis is verified in the
explicit solutions to (\ref{e2.16},~\ref{e2.17}), presented in
references quoted above.
\subsection{\it{Automorphisms, Invariant Description of 3-spaces, and
Quantum Cosmology}} As it is well known, the quantum cosmology
approximation consists in freezing out all but a finite number of
degrees of freedom of the gravitational field and quantize the
rest. This is done by imposing spatial homogeneity. Thus, our --in
principle-- dynamical variables are the scale factors
$\gamma_{\alpha\beta}(t)$, of some spatially homogeneous geometry.

The basic object of the theory, is the wave function $\Psi$, which
must describe the quantum evolution of the 3-geometry.
Consequently, the wave function, will --in principle-- depend on
the 6 $\gamma_{a\beta}$'s. Hence, a question naturally arises;
whether all different $\gamma_{a\beta}$ matrices, are
characterizing different 3-geometries, or not. The answer to this
question, involves the A.I.D.s of the previous section, with the
difference that time does not concern us. Thus, the frozen
analogue of (\ref{e2.9}) will lead us to (\ref{e2.14}) ($\Lambda$
being now, constant) and the integrability condition
(\ref{e2.16}).\\
So, any two matrices $\gamma^{(1)}_{a\beta}$,
$\gamma^{(2)}_{a\beta}$, connected by an element of the
automorphism group $\Lambda^{a}_{\beta}$ (for an arbitrary albeit
given Bianchi Type) i.e. satisfying
$\gamma^{(2)}_{a\beta}=\Lambda^{\mu}_{a}\Lambda^{\nu}_{\beta}\gamma^{(1)}_{\mu\nu}$,
represent the same 3-geometry.

The existence of these A.I.D.'s has very important implications
for the wave functions of a given Bianchi geometry: as we have
proven, points in the configuration space --spanned by the
$\gamma_{a\beta}$'s, named $\Delta$-- that are related through an
automorphism, correspond to spatial line elements that are G.C.T.
related and thus geometrically identifiable. Thus, if we want our
wave-functions to depend only on the Geometry of the three-space
and not on the spatial coordinate system, we must assume them to
be annihilated by the generators of the \underline{entire}
Automorphism Group and not just by the constraint vector fields
$H_{\rho}$, which generate only the so-called inner-automorphisms,
i.e. we have to demand: \be \label{e4.18}
\widehat{X}_{i}\Psi\equiv\lambda^{\rho}_{(i)\mu}\gamma_{\rho\nu}\frac{\partial\Psi}
{\partial\gamma_{\mu\nu}}=0 \ee where:
$\lambda^{\beta}_{(i)a}\equiv(C^{\beta}_{(\rho)a},
\varepsilon^{\beta}_{(i)a})$ are the generators of (the connected
to the identity component of) the Automorphism group and $(i)$
labels the different generators. Depending on the particular
Bianchi Type, the vector fields (in $\Delta$) $X_{(i)}$ may also
include, except of the $H_{\rho}$'s, the generators of the
outer-automorphisms:
$E_{j}\equiv\varepsilon^{\sigma}_{(j)\rho}\gamma_{\sigma\tau}\frac{\partial}{\partial\gamma_{\rho\tau}}$

Using the method of characteristics, the solutions to the set
(\ref{e4.18}), can be found to have the form: \bdm
\Psi=\Psi(q^{i}) \edm where:
\begin{subequations}
\begin{align}
q^{1}(C^{\alpha}_{\mu\nu},
\gamma_{\alpha\beta})&=\frac{m^{a\beta}\gamma_{a\beta}}{\sqrt{\gamma}}\\
q^{2}(C^{\alpha}_{\mu\nu},
\gamma_{\alpha\beta})&=\frac{(m^{a\beta}\gamma_{a\beta})^{2}}{2\gamma}-
\frac{1}{4}C^{a}_{\mu\kappa}C^{\beta}_{\nu\lambda}\gamma_{a\beta}
\gamma^{\mu\nu}\gamma^{\kappa\lambda}\\
q^{3}(C^{\alpha}_{\mu\nu},
\gamma_{\alpha\beta})&=\frac{m}{\sqrt{\gamma}}
\end{align}
\end{subequations}
These three quantities, serve to invariantly describe the geometry
and one can prove the following relevant preposition (T.
Christodoulakis, E. Korfiatis \& G.O. Papadopoulos,
gr-qc/0107050):

\emph{Let $\gamma^{(1)}_{a\beta}, \gamma^{(2)}_{a\beta}, ~\in
\Delta$, and $C^{a}_{\mu\nu}$ be the structure constants of a
given Bianchi Type. If $q^{i}(\gamma^{(1)}, C)=q^{i}(\gamma^{(2)},
C) ~(i=1,2,3)$, then there is $\Lambda$ such that
$\gamma^{(2)}_{a\beta}=\Lambda^{\mu}_{a}\Lambda^{\nu}_{\beta}\gamma^{(1)}_{\mu\nu}$
and $\Lambda ~\in Aut(G)$ i.e.
$C^{\rho}_{\mu\nu}\Lambda^{a}_{\rho}=\Lambda^{\kappa}_{\mu}
\Lambda^{\lambda}_{\nu} C^{a}_{\kappa\lambda}$}.

It is important to notice that the reduction from a 6-dim
configuration space --spanned by the $\gamma_{a\beta}$-- to a
space spanned by the $q$'s, is achieved solely by kinematical
considerations, i.e. the action of G.C.T.'s. Further specification
of the theory, may occur only through the considerations of the
dynamics i.e. the Wheeler-DeWitt equation.

\vspace{1cm} We close this presentation, by giving two examples.
We firstly consider the (see T. Christodoulakis, G. O.
Papadopoulos, Phys. Lett. B \textbf{501} (2001) 264-8):
\subsubsection{\it{Quantization of the most general Bianchi Type
II Vacuum Cosmologies}} In \cite{tchristype2}, we had considered
the quantization of an action corresponding to the most general
Bianchi Type II cosmology, i.e. an action giving Einstein's Field
Equations, derived from the line element: \be
ds^{2}=(N^{2}(t)-N_{a}(t)N^{a}(t))dt^{2}+2N_{a}(t)\sigma^{a}_{i}(x)dx^{i}dt+
\gamma_{\alpha\beta}(t)\sigma^{\alpha}_{i}(x)\sigma^{\beta}_{j}(x)dx^{i}dx^{j}
\ee with: \be
\begin{array}{l}
  \sigma^{a}(x)=\sigma^{\alpha}_{i}(x)dx^{i}\\
  \sigma^{1}(x)=dx^{2}-x^{1}dx^{3} \\
  \sigma^{2}(x)=dx^{3} \\
  \sigma^{3}(x)=dx^{1} \\
  d\sigma^{a}(x)=\frac{1}{2}C^{a}_{\beta\gamma}\sigma^{\beta}\wedge\sigma^{\gamma}\\
  C^{1}_{23}=-C^{1}_{32}=1
\end{array}
\ee
see \cite{ryan}.

As is well known \cite{sneddon}, the Hamiltonian is
$H=\widetilde{N}(t)H_{0}+N^{a}(t)H_{a}$ where:
\be \label{6paper3}
H_{0}=\frac{1}{2}L_{\alpha\beta\mu\nu}\pi^{\alpha\beta}\pi^{\mu\nu}+\gamma
R
\ee
is the quadratic constraint with:
\be
\begin{array}{l}
  L_{\alpha\beta\mu\nu}=\gamma_{\alpha\mu}\gamma_{\beta\nu}+\gamma_{\alpha\nu}\gamma_{\beta\mu}-
\gamma_{\alpha\beta}\gamma_{\mu\nu} \\
  R=C^{\beta}_{\lambda\mu}C^{\alpha}_{\theta\tau}\gamma_{\alpha\beta}\gamma^{\theta\lambda}
\gamma^{\tau\mu}+2C^{\alpha}_{\beta\delta}C^{\delta}_{\nu\alpha}\gamma^{\beta\nu}+
4C^{\mu}_{\mu\nu}C^{\beta}_{\beta\lambda}\gamma^{\nu\lambda}=
C^{\alpha}_{\mu\kappa}C^{\beta}_{\nu\lambda}\gamma_{\alpha\beta}\gamma^{\mu\nu}\gamma^{\kappa\lambda}
\end{array}
\ee
$\gamma$ being the determinant of $\gamma_{\alpha\beta}$ (the last
equality holding only for the Type II case), and:
\be
H_{a}=C^{\mu}_{a\rho}\gamma_{\beta\mu}\pi^{\beta\rho}
\ee
are the linear constraints. Note that $\widetilde{N}$ appearing in
the Hamiltonian, is to be identified with $N/\sqrt{\gamma}$.

The quantities $H_{0}$, $H_{a}$, are weakly vanishing
\cite{dirac}, i.e. $H_{0}\approx 0$, $H_{a}\approx 0$. For all
class A Bianchi Types ($C^{\alpha}_{\alpha\beta}=0$), they can be
seen to obey the following first-class algebra: \be
\begin{array}{l}
  \{H_{0}, H_{0}\}=0 \\
  \{H_{0}, H_{a}\}=0 \\
  \{H_{a}, H_{\beta}\}=-\frac{1}{2}C^{\gamma}_{\alpha\beta}H_{\gamma}
\end{array}
\ee
which ensures their preservation in time i.e. $\dot{H}_{0}\approx
0$, $\dot{H}_{a}\approx 0$ and establishes the consistency of the
action.

If we follow Dirac's general proposal \cite{dirac} for quantizing
this action, we have to turn $H_{0}$, $H_{a}$, into operators
annihilating the wave function $\Psi$.

In the Schr\"{o}dinger representation:
\be
\begin{array}{l}
  \gamma_{\alpha\beta}\rightarrow
\widehat{\gamma}_{\alpha\beta}=\gamma_{\alpha\beta} \\
  \pi^{\alpha\beta}\rightarrow
\widehat{\pi}^{\alpha\beta}=-i\frac{\partial}{\partial\gamma_{\alpha\beta}}
\end{array}
\ee
satisfying the basic Canonical Commutation Relation (CCR)
--corresponding to the classical ones:
\be
[\widehat{\gamma}_{\alpha\beta},
\widehat{\pi}^{\mu\nu}]=-i\delta^{\mu\nu}_{\alpha\beta}=\frac{-i}{2}
(\delta^{\mu}_{\alpha}\delta^{\nu}_{\beta}+\delta^{\mu}_{\beta}\delta^{\nu}_{\alpha})
\ee

The quantum version of the 2 independent linear constraints has
been used to reduce, via the method of characteristics \cite{carabedian},
the dimension of the initial configuration space from 6
($\gamma_{\alpha\beta}$) to 4 (combinations of
$\gamma_{\alpha\beta}$), i.e.
$\Psi=\Psi(q,\gamma,\gamma^{2}_{12}-\gamma_{11}\gamma_{22},\gamma_{12}\gamma_{13}-
\gamma_{11}\gamma_{23})$, where
$q=C^{\alpha}_{\mu\kappa}C^{\beta}_{\nu\lambda}\gamma_{\alpha\beta}\gamma^{\mu\nu}\gamma^{\kappa\lambda}$.

According to Kuha\v{r}'s and Hajicek's \cite{hajiceck} prescription,
the `'kinetic`' part of $H_{0}$ is to be realized as the
conformal Laplacian, corresponding to the reduced metric:
\begin{equation}
L_{\alpha\beta\mu\nu}\frac{\partial x^{i} }{\partial
\gamma_{\alpha\beta} }\frac{\partial x^{j}}{\partial
\gamma_{\mu\nu}}=g^{ij}
\end{equation}
where $x^{i}$, $i=1,2,3,4$, are the arguments of $\Psi$. The
solutions had been presented in \cite{tchristype2}. Note that the
first-class algebra satisfied by $H_{0}$, $H_{a}$, ensures that
indeed, all components of $g^{ij}$ are functions of the $x^{i}$'s.
The signature of the $g^{ij}$, is $(+, +, -, -)$ signaling the
existence of gauge degrees of freedom among the $x^{i}$'s.

Indeed, one can prove \cite{tchriskuchar} that the only gauge
invariant quantity which, uniquely and irreducibly, characterizes
a 3-dimensional geometry admitting Type II symmetry group, is: \be
q=C^{\alpha}_{\mu\kappa}C^{\beta}_{\nu\lambda}\gamma_{\alpha\beta}\gamma^{\mu\nu}\gamma^{\kappa\lambda}
\ee
An outline of the proof, is as follows:\\
Let two hexads $\gamma^{(1)}_{\alpha\beta}$ and
$\gamma^{(2)}_{\alpha\beta}$ be given, such that their
corresponding $q$'s, are equal. Then according to the result given
at the end of the previous section \cite{tchriskuchar}, there
exists an automorphism matrix $\Lambda$ (i.e. satisfying
$C^{a}_{\mu\nu}\Lambda^{\kappa}_{a}=C^{\kappa}_{\rho\sigma}\Lambda^{\rho}_{\mu}\Lambda^{\sigma}_{\nu}$)
connecting them, i.e.
$\gamma^{(1)}_{\alpha\beta}=\Lambda^{\mu}_{\alpha}\gamma^{(2)}_{\mu\nu}\Lambda^{\nu}_{\beta}$.
But as it had been shown in the appendix of \cite{tchristype5},
this kind of changes on $\gamma_{\alpha\beta}$, can be seen to be
induced by spatial diffeomorphisms. Thus, 3-dimensional Type II
geometry, is uniquely characterized by some value of $q$.

Although for full pure gravity, Kucha\v{r} \cite{kuchar} has shown
that there are not other first-class functions, homogeneous and
linear in $\pi^{\alpha\beta}$, except $H_{a}$, imposing the extra
symmetries (Type II), allows for such quantities to exist --as it
will be shown. We are therefore, naturally led to seek the
generators of these extra symmetries --which are expected to chop
off $x^{2}$, $x^{3}$, $x^{4}$. Such quantities are, generally,
called in the literature `'Conditional Symmetries`'.

The automorphism group for Type II, is described by the following
6 generators --in matrix notation and collective form:
\be
\lambda^{a}_{(I)\beta}=\left(
\begin{array}{ccc}
  \kappa+\mu & x & y \\
  0 & \kappa & \rho \\
  0 & \sigma & \mu
\end{array}\right)
\ee
with the property:
\be
C^{a}_{\mu\nu}\lambda^{\kappa}_{a}=C^{\kappa}_{\mu\sigma}\lambda^{\sigma}_{\nu}+C^{\kappa}_{\sigma\nu}\lambda^{\sigma}_{\mu}
\ee
>From these matrices, we can construct the linear --in momenta--
quantities:
\be
A_{(I)}=\lambda^{a}_{(I)\beta}\gamma_{\alpha\rho}\pi^{\rho\beta}
\ee
Two of these, are the $H_{a}$,'s since $C^{a}_{(\rho)\beta}$
correspond to the inner automorphism subgroup --designated by the
x and y parameters, in $\lambda^{a}_{(I)\beta}$. The rest of
them, are the generators of the outer automorphisms and are
described by the matrices:
\be
\varepsilon^{a}_{(I)\beta}=\left(\begin{array}{ccc}
  \kappa+\mu & 0 & 0 \\
  0 & \kappa & \rho \\
  0 & \sigma & \mu
\end{array}\right)
\ee
The corresponding --linear in momenta-- quantities, are:
\be
E_{(I)}=\varepsilon^{a}_{(I)\beta}\gamma_{\alpha\rho}\pi^{\rho\beta}
\ee
The algebra of these --seen as functions on the phase space,
spanned by $\gamma_{\alpha\beta}$ and $\pi^{\mu\nu}$--, is:
\be \label{15paper2}
\begin{array}{l}
  \{E_{I}, E_{J}\}=\widetilde{C}^{K}_{IJ}E_{K} \\
  \{E_{I}, H_{a}\}=-\frac{1}{2}\lambda^{\beta}_{a}H_{\beta} \\
  \{E_{I}, H_{0}\}=-2(\kappa+\mu)\gamma R
\end{array}
\ee
>From the last of  (\ref{15paper2}), we conclude that the subgroup of $E_{I}$'s
with the property $\kappa+\mu=0$, i.e. the traceless generators,
are first-class quantities; their time derivative vanishes. So
let:
\be
\widetilde{E}_{I}=\{E_{I}:~\kappa+\mu=0\}
\ee
Then, the previous statement translates into the form:
\be \label{17paper2}
\dot{\widetilde{E}_{I}}=0 \Rightarrow \widetilde{E}_{I}=c_{I}
\ee
the $c_{I}$'s being arbitrary constants.

Now, these are --in principle-- integrals of motion. Since, as we
have earlier seen, $\widetilde{E}_{I}$'s along with $H_{a}$'s,
generate automorphisms, it is natural to promote the integrals of
motion (\ref{17paper2}), to symmetries --by setting the $c_{I}$'s zero. The
action of the quantum version of these $\widetilde{E}_{I}$'s on
$\Psi$, is taken to be \cite{hajiceck}:
\be
\begin{array}{l}
  \widehat{\widetilde{E}}_{I}\Psi=\varepsilon^{a}_{(I)\beta}\gamma_{\alpha\rho}\frac{\partial\Psi}{\gamma_{\beta\rho}}=0 \\
  \varepsilon^{a}_{(I)a}=0
\end{array} \Bigg\}\Rightarrow \Psi=\Psi(q,\gamma)
\ee

The Wheeler-DeWitt equation now, reads:
\be \label{19paper2}
5q^{2}\frac{\partial^{2} \Psi}{\partial
q^{2}}-3\gamma^{2}\frac{\partial^{2} \Psi}{\partial
\gamma^{2}}+2q\gamma\frac{\partial^{2}\Psi}{\partial
\gamma\partial q}+5q\frac{\partial \Psi}{\partial
q}-3\gamma\frac{\partial \Psi}{\partial \gamma}-2q\gamma\Psi=0
\ee
\textit{Note that:
\be
\nabla^{2}_{c}=\nabla^{2}+\frac{(d-2)}{4(d-1)}R=\nabla^{2}
\ee
since we have a 2-dimensional, flat space, with contravariant
metric:
\be
g^{ij}=\left(\begin{array}{cc}
  5q^{2} & q\gamma \\
  q\gamma & -3\gamma^{2}
\end{array}\right)
\ee
which is Lorentzian}. This equation, can be easily solved by
separation of variables; transforming to new coordinates
$u=q\gamma^{3}$ and $v=q\gamma$, we get the 2 independent
equations:
\be
\begin{array}{l}
  16u^{2}A''(u)+16uA'(u)-cA(u)=0 \\
  B''(v)+\frac{1}{v}B'(v)-(\frac{1}{2v}+\frac{c}{4v^{2}})B(v)=0
\end{array}
\ee where c, is the separation constant. Equation
(\ref{19paper2}), is of hyperbolic type and the resulting wave
function will still not be square integrable. Besides that, the
tracefull generators of the outer automorphisms, are left inactive
--due to the non vanishing CCR with $H_{0}$.

These two facts, lead us to deduce that there must still exist a
gauge symmetry, corresponding to some --would be, linear in
momenta-- first-class quantity. Our starting point in the pursuit
of this, is the third of (\ref{15paper2}). It is clear that we need another
quantity --also linear in momenta-- with an analogous property;
the trace of $\pi^{\mu\nu}$, is such an object. We thus define
the following quantity:
\be
T=E_{I}-(\kappa+\mu)\gamma_{\alpha\beta}\pi^{\alpha\beta}
\ee
in the phase space --spanned by $\gamma_{\alpha\beta}$ and
$\pi^{\mu\nu}$. It holds that:
\be
\begin{array}{l}
  \{T, H_{0}\}=0 \\
  \{T, H_{a}\}=0 \\
  \{T, E_{I}\}=0
\end{array}
\ee
because of:
\be
\begin{array}{l}
  \{E_{I}, \gamma\}=-2(\kappa+\mu)\gamma \\
  \{E_{I}, q\}=0 \\
  \gamma_{\alpha\beta}\{\pi^{\alpha\beta}, q\}=q \\
  \gamma_{\alpha\beta}\{\pi^{\alpha\beta}, \gamma\}=-3\gamma
\end{array}
\ee

Again --as for $\widetilde{E}_{I}$'s--, we see that since $T$, is
first-class, we have that:
\be
\dot{T}=0 \Rightarrow T=const=c_{T}
\ee
another integral of motion. We therefore see, that $T$ has all
the necessary properties to be used in lieu of the tracefull
generator, as a symmetry requirement on $\Psi$. In order to do
that, we ought to set $c_{T}$ zero --exactly as we did with the
$c_{I}$'s, corresponding to $\widetilde{E}_{I}$'s. The quantum
version of $T$, is taken to be:
\be
\widehat{T}=\lambda^{\alpha}_{\beta}\gamma_{\alpha\rho}\frac{\partial}{\partial
\gamma_{\beta\rho}}-(\kappa+\mu)\gamma_{\alpha\beta}\frac{\partial}{\partial
\gamma_{\alpha\beta}}
\ee
Following, Dirac's theory, we require:
\be \label{27paper2}
\widehat{T}\Psi=\lambda^{\alpha}_{\beta}\gamma_{\alpha\rho}\frac{\partial
\Psi}{\partial
\gamma_{\beta\rho}}-(\kappa+\mu)\gamma_{\alpha\beta}\frac{\partial
\Psi}{\partial
\gamma_{\alpha\beta}}=(\kappa+\mu)(q\frac{\partial\Psi}{\partial
q}-\gamma\frac{\partial\Psi}{\partial \gamma})=0
\ee
Equation (\ref{27paper2}), implies that $\Psi(q,\gamma)=\Psi(q\gamma)$ and
thus equation (\ref{19paper2}), finally, reduces to:
\be \label{28paper2}
4w^{2}\Psi''(w)+4w\Psi'(w)-2w\Psi=0
\ee
where, for simplicity, $w\doteq q\gamma$. The solution to this
equation, is:
\be
\Psi=c_{1}I_{0}(\sqrt{2q\gamma})+c_{2}K_{0}(\sqrt{2q\gamma})
\ee
where $I_{0}$ is the modified Bessel function, of the first kind,
and $K_{0}$ is the modified Bessel function, of the second kind,
both with zero argument.

At first sight, it seems that although we have apparently
exhausted the symmetries of the system, we have not yet been able
to obtain a wave function on the space of the 3-geometries, since
$\Psi$ depends on $q\gamma$ and not on $q$ only. On the other
hand, the fact that we have achieved a reduction to one degree of
freedom, must somehow imply that the wave function found must be a
function of the geometry. This puzzle finds its resolution as
follows. Consider the quantity: \be
\Omega=-2\gamma_{\rho\sigma}\pi^{\rho\sigma}+{
\frac{2C^{a}_{\mu\kappa}C^{\beta}_{\nu\lambda}\gamma^{\kappa\lambda}\gamma^{\mu\nu}
\gamma_{\alpha\rho}\gamma_{\beta\sigma}-4C^{\alpha}_{\mu\rho}C^{\beta}_{\nu\sigma}\gamma_{\alpha\beta}\gamma^{\mu\nu}}{q}}\pi^{\rho\sigma}
\ee This can also be seen to be first-class, i.e. \be
\dot{\Omega}=0 \Rightarrow \Omega=const=c_{\Omega} \ee Moreover,
it is a linear combination of $T$, $\widetilde{E}_{I}$'s, and
$H_{a}$'s, and thus $c_{\Omega}=0$. Now it can be verified that
$\Omega$, is nothing but: \be
\frac{1}{N(t)}(\frac{\dot{\gamma}}{\gamma}+\frac{1}{3}\frac{\dot{q}}{q})
\ee So: \be \gamma q^{1/3}=\vartheta=constant \ee Without any loss
of generality, and since $\vartheta$ is not an essential constant
of the classical system (see \cite{tchrisaut} and reference [18]
therein), we set $\vartheta=1$. Therefore: \be
\Psi=c_{1}I_{0}(\sqrt{2}q^{1/3})+c_{2}K_{0}(\sqrt{2}q^{1/3}) \ee
where $I_{0}$ is the modified Bessel function, of the first kind,
and $K_{0}$ is the modified Bessel function, of the second kind,
both with zero argument.

As for the measure, it is commonly accepted that, there is not a
unique solution. A natural choice, is to adopt the measure that
makes the operator in (\ref{28paper2}), hermitian --that is:
\be
\mu(q)\propto q^{-1}
\ee
It is easy to find combinations of $c_{1}$ and $c_{2}$ so that
the probability $\mu(q)|\Psi|^{2}$, be defined.

Note that putting the constant associated with $\Omega$, equal to
zero, amounts in restricting to a subset of the classical
solutions, since $c_{\Omega}$, is one of the two essential
constants of Taub's solution. One could keep that constant, at the
expense of arriving at a wave function with explicit time
dependence, since then: \be
\gamma=q^{-1/3}Exp[\int{c_{\Omega}N(t)}dt] \ee We however,
consider more appropriate to set that constant zero, thus arriving
at a $\Psi$ depending on $q$ only, and decree its applicability to
the entire space of the classical solutions. Anyway this is not
such a blunder, since $\Psi$ is to give weight to all states,
--being classical ones, or not.

\vspace{1cm} And the last example  (see T. Christodoulakis, G.
Gakis \& G. O. Papadopoulos, gr-qc/0106065):
\subsubsection{\it{Conditional Symmetries and the Quantization of Bianchi Type I Vacuum Cosmologies with
Cosmological Constant}}

\emph{Note: The original work, deals with both the cases; the
models with a vanishing and those with non vanishing cosmological
constant.}

The case of Bianchi Type I geometries, has been repeatedly treated
in the literature --both at the classical level \cite{Landaupaper}
and the quantum level \cite{Ashtekarpaper}. The main reason, is
the simplicity brought by the vanishing structure constants, i.e.
the high spatial symmetry of the model. Thus, the most general of
these models, is described by the 6 scale factors
$\gamma_{\alpha\beta}(t)$ and the lapse function $N(t)$ --the
shift vector $N^{\alpha}(t)$, being absent due to the non
existence of the $H_{a}$'s (linear constraints). The absence of
$H_{a}$s presents --at first sight-- the complication that no
reduction of the initial configuration space, is possible --in
contrast to what happens in other Bianchi Types \cite{Chrispaper}.

In what follows, we present a complete reduction of the initial
configuration space for Bianchi Type I geometry, when the
cosmological constant is present. A wave function, which depends
on one degree of freedom, is found.

As is well known (first of \cite{sneddon}) the Hamiltonian of the
above system is: \bdm
H=\widetilde{N}(t)H_{0}+N^{\alpha}(t)H_{\alpha}\edm where: \be
\label{hamiltonian}
H_{0}=\frac{1}{2}L_{\alpha\beta\mu\nu}\pi^{\alpha\beta}\pi^{\mu\nu}+\gamma\Lambda\ee

Thus, the only operator which must annihilate the wave function,
is $\widehat{H}_{0}$; and the Wheeler-DeWitt equation
$\widehat{H}_{0}\Psi=0$, will produce a wave function, initially
residing on a 6-dimensional configuration space --spanned by
$\gamma_{\alpha\beta}$'s. The discussion however, does not end
here. If the linear constraints existed, a first reduction of the
initial configuration space, would take place \cite{hajiceck}. New
variables, instead of the 6 scale factors, would emerge --say
$q^{i}$, with $i<6$. Then a new ''physical'' metric would be
induced: \be \label{physicalmetric}
g^{ij}=L_{\alpha\beta\mu\nu}\frac{\partial q^{i} }{\partial
\gamma_{\alpha\beta}}\frac{\partial q^{j}}{\partial
\gamma_{\mu\nu}} \ee According to Kucha\v{r}'s and Hajicek's
\cite{hajiceck} prescription, the ''kinetic'' part of $H_{0}$
would have to be realized as the conformal Laplacian (in order for
the equation to respect the conformal covariance of the classical
action), based on the physical metric (\ref{physicalmetric}). In
the presence of conditional symmetries, further reduction can take
place, a new physical metric would then be defined similarly, and
the above mentioned prescription, would have to be used after the
final reduction \cite{Chrispaper,kuchar}.

The case of Bianchi Type I, is an extreme example in which all the
linear constraints, vanish identically; thus no initial physical
metric, exists --another peculiarity reflecting the high spatial
symmetry of the model  under consideration. In compensation, a lot
of integrals of motion exist and the problem of reduction, finds
its solution through the notion of \emph{`'Conditional
Symmetries`'}.

The automorphism algebra of this Type, has been exhaustively
treated in the literature --see e.g. \cite{harvey}. The relevant
group, is that of the constant, real, $3\times 3$, invertible,
matrices i.e. $GL(3,\Re)$. The generators of this automorphism
group, are (in a collective form and matrix notation) the
following 9 --one for each parameter: \be \label{Lambdas}
\lambda^{\alpha}_{(I)\beta}=\left(
\begin{array}{ccc}
  a  & \beta & \delta \\
  \epsilon & \zeta & \eta \\
  \theta & \sigma & \rho
\end{array}\right),~~~ I \in [1,\ldots,9]
\ee with the defining property: \be
C^{\alpha}_{\mu\nu}\lambda^{\kappa}_{\alpha}=C^{\kappa}_{\mu\sigma}\lambda^{\sigma}_{\nu}+C^{\kappa}_{\sigma\nu}\lambda^{\sigma}_{\mu}.
\ee Exponentiating all these matrices, one obtains the outer
automorphism group of Type I.

For full pure gravity, Kucha\v{r} \cite{kuchar} has shown that
there are no other first-class functions, homogeneous and linear
in the momenta, except the linear constraints. If however, we
impose extra symmetries (e.g. the Bianchi Type I --here
considered), such quantities may emerge --as it will be shown. We
are therefore --according to Dirac \cite{dirac}-- justified to
seek the generators of these extra symmetries, whose
quantum-operator analogues will be imposed as additional
conditions on the wave function. Thus, these symmetries are
expected to lead us to the final reduction, by revealing the true
degrees of freedom. Such quantities are, generally, called in the
literature \emph{`'Conditional Symmetries`'} \cite{kuchar}. From
matrices (\ref{Lambdas}), we can construct the linear --in
momenta-- quantities: \be \label{epsilons}
E_{(I)}=\lambda^{\alpha}_{(I)\beta}\gamma_{\alpha\rho}\pi^{\rho\beta}
\ee

In order to write analytically these quantities, the following
base is chosen: \be \label{basis}
\begin{array}{ccc}
 \lambda_{1}= \left(\begin{array}{ccc}
  0 & 1 & 0 \\
  0 & 0 & 0 \\
  0 & 0 & 0
\end{array}\right), & \lambda_{2}=\left(\begin{array}{ccc}
  0 & 0 & 1 \\
  0 & 0 & 0 \\
  0 & 0 & 0
\end{array}\right), & \lambda_{3}=\left(\begin{array}{ccc}
  0 & 0 & 0 \\
  0 & 0 & 1 \\
  0 & 0 & 0
\end{array}\right) \\
  \lambda_{4}=\left(\begin{array}{ccc}
  0 & 0 & 0 \\
  0 & 0 & 0 \\
  0 & 1 & 0
\end{array}\right), & \lambda_{5}=\left(\begin{array}{ccc}
  0 & 0 & 0 \\
  0 & 0 & 0 \\
  1 & 0 & 0
\end{array}\right), & \lambda_{6}=\left(\begin{array}{ccc}
  0 & 0 & 0 \\
  1 & 0 & 0 \\
  0 & 0 & 0
\end{array}\right) \\
  \lambda_{7}=\left(\begin{array}{ccc}
  1 & 0 & 0 \\
  0 & -1 & 0 \\
  0 & 0 & 0
\end{array}\right), & \lambda_{8}=\left(\begin{array}{ccc}
  0 & 0 & 0 \\
  0 & -1 & 0 \\
  0 & 0 & 1
\end{array}\right), & \lambda_{9}=\left(\begin{array}{ccc}
  1 & 0 & 0 \\
  0 & 1 & 0 \\
  0 & 0 & 1
\end{array}\right)
\end{array}
\ee

It is straightforward to calculate the Poisson Brackets of
$E_{(I)}$ with $H_{0}$: \be \label{commutatorEH} \{E_{(I)},
H_{0}\}=-\gamma\Lambda\lambda^{a}_{a} \ee But, it holds that: \be
\dot{E}_{(I)}=\{E_{(I)},H_{0}\}=-\gamma\Lambda\lambda^{a}_{a}\ee
--the last equality emerging by virtue of (\ref{commutatorEH}).
Thus: \be \label{integralsofmotion}
\dot{E}_{(I)}=\{E_{(I)},H_{0}\}=0 \Rightarrow
E_{(I)}=K_{(I)}=\textrm{constants},~~~ I \in [1,\ldots,8] \ee We
therefore conclude that, \underline{the first eight quantities}
$E_{(I)}$, are first-class, and thus integrals of motion. Out of
the eight quantities $E_{(I)}$, only five are functionally
independent (i.e. linearly independent, if we allow for the
coefficients of the linear combination, to be functions of the
$\gamma_{\alpha\beta}$ s); numerically, they are all independent.

The algebra of $E_{(I)}$ can be easily seen to be: \be
\label{algebraofepsilons}
\{E_{(I)},E_{(J)}\}=-\frac{1}{2}C^{M}_{IJ}E_{(M)},~~~I,J,M \in
[1,\ldots,9] \ee where: \be \label{algebraoflambda}
[\lambda_{(I)},\lambda_{(J)}]=C^{M}_{IJ}\lambda_{(M)},~~~I,J,M \in
[1,\ldots,9] \ee the square brackets denoting matrix
commutation.\\
The non vanishing structure constants of the algebra
(\ref{algebraoflambda}), are found to be: \be \label{liealgebra}
\begin{array}{lllll}
  C^{2}_{13}=1 & C^{4}_{15}=-1 & C^{7}_{16}=1 & C^{1}_{18}=-1 & C^{1}_{17}=-2 \\
  C^{1}_{24}=1 & C^{7}_{25}=1 & C^{8}_{25}=-1 & C^{3}_{26}=-1 & C^{2}_{27}=-1 \\
  C^{2}_{28}=1 & C^{8}_{34}=-1 & C^{6}_{35}=1 & C^{3}_{37}=1 & C^{3}_{38}=2 \\
  C^{5}_{46}=1 & C^{4}_{47}=-1 & C^{4}_{48}=-2 & C^{5}_{57}=1 & C^{5}_{58}=-1 \\
  C^{6}_{67}=2 & C^{6}_{68}=1
\end{array}
\ee

At this point, in order to achieve the desired reduction, we
propose that the quantities $E_{(I)}$ --with $I \in
[1,\ldots,8]$-- must be promoted to operational conditions acting
on the requested wave function $\Psi$ --since they are first class
quantities and thus integrals of motion (see
(\ref{integralsofmotion})). In the Schr\"{o}dinger representation:
\be \label{epsilonuponPsi}
\widehat{E}_{(I)}\Psi=-i\lambda^{\tau}_{(I)\alpha}\gamma_{\tau\beta}\frac{\partial
\Psi}{\partial \gamma_{\alpha\beta}}=K_{(I)}\Psi,~~~I \in
[1,\ldots,8] \ee In general, systems of equations of this type,
must satisfy some consistency conditions (i.e. the Frobenius
Theorem): \be
\begin{array}{ccc}
  \widehat{E}_{(J)}\Psi=K_{(J)}\Psi & \Rightarrow & \widehat{E}_{(I)}\widehat{E}_{(I)}\Psi=K_{(I)}K_{(J)}\Psi\\
  \widehat{E}_{(I)}\Psi=K_{(I)}\Psi & \Rightarrow & \widehat{E}_{(J)}\widehat{E}_{(I)}\Psi=K_{(J)}K_{(I)}\Psi
\end{array}
\ee Subtraction of these two and usage of
(\ref{algebraofepsilons}), results in: \be \label{selectionrule}
K^{M}_{IJ}\widehat{E}_{(M)}\Psi=0 \Rightarrow
C^{M}_{IJ}K_{(M)}=0\ee i.e. a selection rule for the numerical
values of the integrals of motion. Consistency conditions
(\ref{selectionrule}) and the Lie Algebra (\ref{liealgebra}),
impose that $K_{1}=\ldots=K_{8}=0$. If we also had $E_{(9)}$ (as
is the case when $\Lambda=0$) then $K_{9}$ would remain arbitrary.
With this outcome, and using the method of characteristics, the
system of the five functionally independent P.D.E. s
(\ref{epsilonuponPsi}), can be integrated. The result is:
 \be \label{wavefunction}
\Psi=\Psi(\gamma) \ee i.e. an arbitrary (but well behaved)
function of $\gamma$ --the determinant of the scale factor matrix.

\emph{A note is pertinent here; from basic abstract algebra, is
well known that the basis of a linear vector space, is unique
--modulo linear mixtures. Thus, although the form of the system
(\ref{epsilonuponPsi}) is base dependent, its solution
(\ref{wavefunction}), is base independent.}

The next step, is to construct the Wheeler-DeWitt equation which
is to be solved by the wave function (\ref{wavefunction}). The
degree of freedom, is 1; the $q=\gamma$. According to Kucha\v{r}'s
proposal \cite{hajiceck}, upon quantization, the kinetic part of
Hamiltonian is to be realized as the conformal Beltrami operator
-- based on the induced physical metric --according to
(\ref{physicalmetric}), with $q=\gamma$: \be g^{11}=
L_{\alpha\beta\mu\nu}\frac{\partial \gamma }{\partial
\gamma_{\alpha\beta}}\frac{\partial \gamma}{\partial
\gamma_{\mu\nu}}=L_{\alpha\beta\mu\nu}\gamma^{2}\gamma^{\alpha\beta}\gamma^{\mu\nu}
=-3\gamma^{2}\ee In the Schr\"{o}dinger representation: \be
\frac{1}{2}L_{\alpha\beta\mu\nu}\pi^{\alpha\beta}\pi^{\mu\nu}
\rightarrow -\frac{1}{2}\Box^{2}_{c} \ee where: \be \label{box}
\Box^{2}_{c}=\Box^{2}=\frac{1}{\sqrt{g_{11}}}~\partial_{\gamma}
\{\sqrt{g_{11}}~g^{11}~\partial_{\gamma}\} \ee is the
1--dimensional Laplacian based on $g_{11}$ ($g^{11}g_{11}=1$).
Note that in 1--dimension the conformal group is totally contained
in the G.C.T. group, in the sense that any conformal
transformation of the metric can not produce any change in the
--trivial-- geometry and is thus reachable by some G.C.T.
Therefore, no extra term in needed in (\ref{box}), as it can also
formally be seen by taking the limit $d=1,~R=0$ in the general
definition: \bdm
\Box^{2}_{c}\equiv\Box^{2}+\frac{(d-2)}{4(d-1)}R=\Box^{2} \edm
Thus: \be H_{0}\rightarrow
\widehat{H}_{0}=-\frac{1}{2}(-3\gamma^{2}\frac{\partial^{2}}{\partial\gamma}
-3\gamma\frac{\partial}{\partial\gamma})+\Lambda\gamma \ee So, the
Wheeler-DeWitt equation --by virtue of (\ref{wavefunction})--,
reads: \be
\widehat{H}_{0}\Psi=\gamma^{2}\Psi''+\gamma\Psi'+\frac{2}{3}\gamma\Lambda\Psi=0
\ee The general solution to this equation, is: \be
\Psi(\gamma)=c_{1}J_{0}(2\sqrt{\frac{2\gamma\Lambda}{3}})
+c_{2}Y_{0}(2\sqrt{\frac{2\gamma\Lambda}{3}}) \ee where $J_{n}$
and $Y_{n}$, are the Bessel Functions of the first and second kind
respectively --both with zero argument-- and $c_{1},c_{2}$,
arbitrary constants.

An important element for selecting the measure, is the conformal
covariance; the supermetric $L^{\alpha\beta\mu\nu}$ is known only
up to rescalings, because instead of $\widetilde{N}(t)$ one can
take any $\overline{N}(t)=\widetilde{N}(t)e^{-2\omega}$ (with
$\omega=\omega(\gamma_{\alpha\beta})$) and consequently
$\overline{L}^{\alpha\beta\mu\nu}(t)=L^{\alpha\beta\mu\nu}(t)e^{2\omega}$.
This property, is also inherited to the physical metric
(\ref{physicalmetric}) and is the reason for the Kucha\v{r}'s
recipe, adopted in this work.

It is natural that the proposed measure $\mu$, must be such that
the probability density $\mu\mid\Psi\mid^{2}$, be invariant under
these scalings. Recalling that $\overline{\Psi}=\Psi
e^{(2-\mathcal{D})\omega/2}$, we conclude that $\mu$ must scale as
$\overline{\mu}=\mu e^{(\mathcal{D}-2)\omega}$. The natural
measure under which the Wheeler-DeWitt operator is hermitian, is
$\sqrt{\textrm{Det(physical metric)}}$, but it scales as
$\sqrt{\overline{\textrm{Det(physical
metric)}}}=\sqrt{\textrm{Det(physical
metric)}}e^{\mathcal{D}\omega}$.

We are thus after a quantity $\xi$ --preferably constant, (so that
the hermiticity is preserved)-- which scales as
$\overline{\xi}=\xi e^{-2\omega}$. It is not difficult to imagine
such a quantity: The inverse of any product of
$E_{(I)\alpha\beta}$ with $E_{(J)\mu\nu}$ (where
$E_{(I)\alpha\beta}=1/2(\lambda^{\kappa}_{a}\gamma_{\kappa\beta}+(\alpha\leftrightarrow
\beta))$) has the desired property. Indeed the $E_{(I)}$ s do not
scale at all, while the supermetric scales as mentioned before.
The group metric $\Theta_{IJ}=C^{F}_{IS}C^{S}_{JF}$ can serve to
close the group indices of $E_{(I)\alpha\beta}$. So, we arrive at
the quantity: \be \label{xi}
\xi=\frac{1}{L^{\alpha\beta\mu\nu}\Theta^{IJ}E_{(I)\alpha\beta}E_{(J)\mu\nu}}
\ee (where $\Theta^{IJ}$ is the inverse of the group metric)
having the desired property and being also a constant. Using the
Lie algebra (\ref{liealgebra}), one obtains: \be \Theta_{IJ}\equiv
C^{F}_{IS}C^{S}_{JF}=\left(\begin{array}{cccccccc}
  0 & 0 & 0 & 0 & 0 & 6 & 0 & 0  \\
  0 & 0 & 0 & 0 & 6 & 0 & 0 & 0  \\
  0 & 0 & 0 & 6 & 0 & 0 & 0 & 0  \\
  0 & 0 & 6 & 0 & 0 & 0 & 0 & 0  \\
  0 & 6 & 0 & 0 & 0 & 0 & 0 & 0  \\
  6 & 0 & 0 & 0 & 0 & 0 & 0 & 0  \\
  0 & 0 & 0 & 0 & 0 & 0 & 6 & 12 \\
  0 & 0 & 0 & 0 & 0 & 0 & 12 & 6
\end{array}\right) \ee
Thus: \be \Theta^{IJ}=\left(\begin{array}{cccccccc}
  0 & 0 & 0 & 0 & 0 & 1/6 & 0 & 0  \\
  0 & 0 & 0 & 0 & 1/6 & 0 & 0 & 0  \\
  0 & 0 & 0 & 1/6 & 0 & 0 & 0 & 0  \\
  0 & 0 & 1/6 & 0 & 0 & 0 & 0 & 0  \\
  0 & 1/6 & 0 & 0 & 0 & 0 & 0 & 0  \\
  1/6 & 0 & 0 & 0 & 0 & 0 & 0 & 0  \\
  0 & 0 & 0 & 0 & 0 & 0 & 1/9 & -1/18  \\
  0 & 0 & 0 & 0 & 0 & 0 & -1/18 & 1/9
\end{array}\right) \ee
After a straightforward calculation, one finds that: \be
\xi=\frac{12}{5} \ee

The product of $\xi$ with the respective natural measure, defines
the final expression for the measure $\mu$.

It is fair to say that the problem of selection of the `'correct`'
measure, is not yet resolved; it is a reflection of the problem of
time in Quantum Gravity \cite{isham}.

Another issue that has not been touched upon is the problem of
selecting a unique wave faction. In the path integral approach, to
quantum cosmology, there is the Hartle-Hawking `'no boundary
proposal`' \cite{hh}. In the canonical approach, there are various
forms of the Vilenkin proposal \cite{vile}.

Finally, the problem of decoherence (i.e. of reconstruction of
classical trajectories, from the knowledge of the wave function),
has occupied several workers in the field \cite{laf}.

\end{document}